%% file: KZ6.tex
\documentclass[conference]{IEEEtran}
\IEEEoverridecommandlockouts
\include{micros}

\begin{document}

\title{A  Modified KZ Reduction Algorithm}

\author{\IEEEauthorblockN{Jinming~Wen ~\textit{Student Member}\emph{,~IEEE}}
\IEEEauthorblockA{Department of Mathematics and Statistics\\
McGill University, Montreal, Canada\\
Email: jwen@math.mcgill.ca}% This "%" stops a space
\and
\IEEEauthorblockN{Xiao-Wen~Chang}
\IEEEauthorblockA{School of Computer Science\\
McGill University, Montreal, Canada\\
Email: chang@cs.mcgill.ca}
\thanks{This work was supported by NSERC of Canada grant  217191-12.}}

%\markboth{}%
%{Shell \MakeLowercase{\textit{et al.}}: Bare Demo of IEEEtran.cls for Journals}


\maketitle

\begin{abstract}
The Korkine-Zolotareff (KZ) reduction has been  used in communications and cryptography.
In this paper, we modify a very recent KZ reduction algorithm proposed by  Zhang et al.,
resulting in a new algorithm, which can be much faster and more numerically reliable,
especially when the basis matrix is ill conditioned.
\end{abstract}



% \fi
\IEEEpeerreviewmaketitle

\section{Introduction}
For any full column rank matrix $\A\in \mathbb{R}^{m\times n}$, the lattice $\mathcal{L}(\A)$ generated by $\A$ is defined by
\beq
\label{e:latticeA}
\mathcal{L}(\A)=\{\A\z|\z \in \mathbb{Z}^n\}.
\eeq
The columns of $\A$  form a basis of $\mathcal{L}(\A)$.
For any $n\geq2$, $\mathcal{L}(\A)$ has infinity many  bases and any of two are connected by a unimodular matrix $\Z$,
i.e.,  $\Z \in \mathbb{Z}^{n\times n}$ and $\det(\Z)=\pm1$.  Specifically, for each given lattice basis matrix
$\A\in \mathbb{R}^{m\times n}$, $\A\Z$ is also a basis matrix of $\mathcal{L}(\A)$ if and only if $\Z$ is unimodular,
see, e.g., \cite{AgrEVZ02}.

The process of selecting a good basis for a given lattice, given some criterion,
is called lattice reduction.
In many applications, it is advantageous if the basis vectors are short
and close to be orthogonal \cite{AgrEVZ02}.
For more than a century, lattice reduction have been investigated by many people and several  types of reductions
have been proposed,  including the KZ reduction \cite{KZ73}, the Minkowski reduction \cite{Min96},
the LLL reduction \cite{LenLL82} and Seysen's reduction \cite{Sey93} etc.

%Lattice reduction plays an important role in many research areas, such as, in combinatorial optimization (see, e.g., \cite{Eis10}),
%GPS (see, e.g., \cite{HasB98}), cryptography (see, e.g., \cite{MicR08, HanS07, HanPS11}),
%communications (see, e.g., \cite{Mow94, AgrEVZ02, DamGC03}), number theory (see, e.g., \cite{GolRS00, GurSS00}),
%and transportation science (see, e.g., \cite{KuuSS14}), etc.
%For more details, see the survey paper \cite{WubSJM11} and references therein.

Lattice reduction plays an important role in many research areas, such as,
%in combinatorial optimization (see, e.g., \cite{Eis10}),
cryptography (see, e.g., \cite{HanPS11}),
communications (see, e.g., \cite{AgrEVZ02, WubSJM11}) and GPS (see, e.g., \cite{Teu96}),
where the closest vector problem (CVP) and/or the shortest vector problem (SVP) need to be solved:
\beq
\label{e:ILS}
\min_{\x\in\mathbb{Z}^n}\|\y-\A\x\|_2^2,
\eeq
\beq
\label{e:SVP}
\min_{\x \in\mathbb{Z}^n\backslash \{\0\}} \|\A\x\|_2^2.
\eeq

The often used lattice reduction is the LLL reduction, which can be computed in polynomial time under some conditions
and has some nice properties, see, e.g., \cite{ChaWX13} for some latest results.
In some communication applications, one needs to solve a sequence of CVPs, where $\y$'s are different, but $\A$'s are identical.
In this case, instead of using the LLL reduction, one usually uses the
KZ reduction \cite{KZ73} to do reduction, since sphere decoding for solving these CVPs becomes
more efficient, although the KZ reduction costs more than the LLL reduction.

%The properties of the KZ reduced bases has been studied in \cite{LagLS90}.
%As computing the KZ reduced matrix is too time-consuming,
%the blockwise KZ (BKZ) reduction, which outputs a BKZ-reduced matrix with each local block matrix being KZ reduced,
%has been proposed in \cite{SchE94}.
%For more details and its variant, see, e.g., \cite{CheN11}.

There are various KZ reduction algorithms,
%see, e.g., \cite{Kan83}, \cite{Hel85}, \cite{Kan87}, \cite{Sch87}, \cite{AgrEVZ02}.
see, e.g.,  \cite{Hel85}, \cite{Kan87}, \cite{Sch87}, \cite{AgrEVZ02}.
Very recently, another KZ reduction algorithm was proposed in \cite{ZhaQW12}.
Like in \cite{AgrEVZ02},  the LLL-aided Schnorr-Euchner search strategy \cite{SchE94} is used to solve the $n-1$ SVPs in \cite{ZhaQW12}.
But instead of using   Kannan's basis expansion method used in \cite{Kan87} and \cite{AgrEVZ02},
it uses a new basis expansion method which is more efficient.

In this paper, we will propose a new KZ reduction algorithm,
which improves the basis expansion method proposed in \cite{ZhaQW12}.
Like \cite{ZhaQW12}, we assume floating point arithmetic with fixed precision is used in the computation.
Numerical results indicate that the modified algorithm  can be much faster and more  numerically reliable.
%Simulation results will be presented to illustrate this.

The rest of the paper is organized as follows.
In section \ref{s:reduction}, we introduce the LLL and KZ reductions.
In section \ref{s:KZ}, we introduce our modified KZ reduction algorithm.
Some simulation results are given in section \ref{s:sim} to show the efficiency and numerical reliability of our new algorithm.
Finally, we summarize this paper in section \ref{s:sum}.

In this paper, %let $\mathbb{R}^{m\times n}$ and $\mathbb{Z}^{m\times n}$ be the space of the $m\times n$
%real matrices and integer matrices, respectively.
boldface lowercase letters denote column vectors and boldface uppercase letters denote matrices.
For a matrix $\A$, let $a_{ij}$ be its $(i,j)$ element and
$\A_{i:j,k:\ell}$ be the submatrix containing elements with row indices from $i$ to $j$ and column indices from $k$ to $\ell$.
Denote $\e_1=[1,0,\ldots, 0]^T$, whose dimension depends on the context.

\section{LLL and KZ Reductions}\label{s:reduction}

Assume that $\A$ in \eqref{e:latticeA} has the QR factorization
\beq
\label{e:QR}
\A=[\Q_1, \Q_2]\bmx\R \\ \0 \emx
\eeq
where $[\underset{n}{\Q_1}, \underset{m-n}{\Q_2}]\in \Rmbm$ is orthogonal and $\R\in \Rnbn$ is upper triangular.
%With loss of generality, we assume that the diagonal entries of upper triangular matrices in this paper are positive.

After the QR factorization of $\A$, the LLL reduction \cite{LenLL82} reduces the matrix $\R$ in \eqref{e:QR} to $\bbR$
through the QRZ factorization:
\beq
\label{e:QRZ}
\bbQ^T \R \Z = \bbR,
\eeq
where $\bbQ  \in \mathbb{R}^{n\times n}$ is orthogonal,
$\Z\in   \mathbb{Z}^{n\times n}$ is  unimodular  and
$\bbR\in \mathbb{R}^{n\times n}$ is upper triangular and  satisfies the following conditions:
\begin{align}
&|\br_{ik}|\leq\frac{1}{2} |\br_{ii}|, \quad i=1, 2, \ldots, k-1 \label{e:criteria1} \\
&\delta\, \br_{k-1,k-1}^2 \leq   \br_{k-1,k}^2+ \br_{kk}^2,\quad k=2, 3, \ldots, n \label{e:criteria2}
\end{align}
where $\delta$ is a constant satisfying $1/4 < \delta \leq 1$.
The matrix $\A\Z$ is said to be  LLL reduced.
Equations \eqref{e:criteria1} and  \eqref{e:criteria2} are referred to as the size-reduced condition and the Lov\'{a}sz condition, respectively.

Similarly, after the QR factorization of $\A$, the KZ reduction reduces the matrix $\R$ in \eqref{e:QR} to $\bbR$ in \eqref{e:QRZ},
where $\bbR$ satisfies \eqref{e:criteria1} and
\begin{align}
\label{e:criteria2KZ}
%|\br_{ii}|\leq \lambda_1(\bbR_{i:n,i:n}), \quad i=1, 2, \ldots, n,
| \br_{ii}| =\min_{\x\,\in\,\mathbb{Z}^{n-i+1}\backslash \{\0\}}\|\bbR_{i:n,i:n}\x\|_2, \ \
i=1,\ldots, n.
\end{align}
The matrix $\A\Z$ is said to be  KZ reduced.
Note that if a matrix is KZ reduced, it must be LLL reduced for $\delta=1$.

\section{A modified KZ reduction algorithm }\label{s:KZ}
In this section, we first introduce the KZ reduction algorithm given in \cite{ZhaQW12},
then propose a modified algorithm.

\subsection{The KZ Reduction Algorithm in \cite{ZhaQW12}}

From the definition of the KZ reduction, the reduced matrix $\bbR$ satisfies both \eqref{e:criteria1} and \eqref{e:criteria2KZ}.
If the QRZ factorization in  \eqref{e:QRZ} gives $\bbR$  satisfying \eqref{e:criteria2KZ},
then we can easily apply size reductions to $\bbR$ such that \eqref{e:criteria1} holds.
Thus, in the following, we will only show how to obtain $\bbR$ such that \eqref{e:criteria2KZ} holds.

The algorithm needs $n-1$ steps.
Suppose that at the end of step $k-1$,  one has found an orthogonal matrix
$\Q^{(k-1)} \in \Rbb^{n\times n}$, a unimodular matrix $\Z^{(k-1)}\in \Zbb^{n\times n}$ and
an upper triangular $\R^{(k-1)}\in \Rbb^{n\times n}$ such that
%$$r^{(k-1)}_{11}=\br_{11}, \ldots, r^{(k-1)}_{k-1,k-1}=\br_{k-1,k-1}$$
\begin{align}
\label{e:recursionk-1}
(\Q^{(k-1)})^T\R\Z^{(k-1)}=\R^{(k-1)}
\end{align}
where for $ i=1,\ldots,  k-1$,
\beq
|r^{(k-1)}_{ii}|= \min_{\x\,\in\,\mathbb{Z}^{n-i+1}\backslash \{\0\}}\| \R^{(k-1)}_{i:n,i:n} \x\|_2. \ \  \label{e:diagk-1}
\eeq
%Then, the KZ reduction algorithm in \cite{ZhaQW12} determines $\br_{kk}$ as follows.

At step $k$, like \cite{AgrEVZ02},  \cite{ZhaQW12} uses the LLL-aided Schnorr-Euchner search strategy \cite{SchE94}  to solve
the SVP:
\begin{align}
\label{e:SVPk}
\x^{(k)}=\arg \min_{\x\,\in \mathbb{Z}^{n-k+1}\setminus \{\0\}}\|\R^{(k-1)}_{k:n,k:n}\x\|_2^2.
\end{align}
Then, unlike other KZ reduction algorithms,  \cite{ZhaQW12} finds the unimodular matrix by
expanding $\R^{(k-1)}_{k:n,k:n}\x^{(k)}$ to
a basis for the lattice $\{\R^{(k-1)}_{k:n,k:n}\x: \x\in \mathbb{Z}^{n-k+1}\}$.
Specifically,  \cite{ZhaQW12} first  constructs a  unimodular matrix $\widetilde{\Z}^{(k)}\in \Zbb^{(n-k+1)\times (n-k+1)}$ whose first column is $\x^{(k)}$,
i.e.,
\beq
\widetilde{\Z}^{(k)} \e_1 = \x^{(k)}
\label{e:zkxk}
\eeq
and then finds an orthogonal matrix $\widetilde{\Q}^{(k)}$ to bring $\R^{(k-1)}_{k:n,k:n}\widetilde{\Z}^{(k)}$
back to an upper triangular matrix $\widetilde{\R}^{(k)}$, i.e.,
\begin{align}
\label{e:recursionk-12}
(\widetilde{\Q}^{(k)})^T\R^{(k-1)}_{k:n,k:n}\widetilde{\Z}^{(k)}=\widetilde{\R}^{(k)}.
\end{align}

Based on \eqref{e:recursionk-1} and \eqref{e:recursionk-12}, we define
\begin{align}
\label{e:Qk}
\Q^{(k)}&=\Q^{(k-1)}\bsmx \I_{k-1}&\0\\ \0&\widetilde{\Q}^{(k)}\esmx,\\
\label{e:Rk}
\R^{(k)}&=\bsmx \R^{(k-1)}_{1:k-1, 1:k-1}&\R^{(k-1)}_{1:k-1, k:n}\widetilde{\Z}^{(k)}\\ \0&\widetilde{\R}^{(k)}\esmx,\\
\Z^{(k)}&=\Z^{(k-1)}\bsmx \I_{k-1}&\0\\ \0&\widetilde{\Z}^{(k)}\esmx.
\label{e:Zk}
\end{align}
Here $\Q^{(k)}$ is orthogonal, $\R^{(k)}$ is upper triangular and $\Z^{(k)}$ is unimodular.
Then,  combining \eqref{e:recursionk-1} and \eqref{e:recursionk-12}, we obtain
\begin{align}
\label{e:recursionk}
(\Q^{(k)})^T\R\Z^{(k)}=\R^{(k)}.
\end{align}

At the end of step $n-1$,  we get $\R^{(n-1)}$, which is just $\bbR$ in \eqref{e:QRZ}.
In the following we explain why \eqref{e:criteria2KZ} holds.

From \eqref{e:Rk} and  \eqref{e:recursionk-12},  it is easy to verify that for $i=1, \ldots, k$,
\begin{align} \label{e:rink}
\R_{i:n,i:n}^{(k)} = \bsmx \I_{k-i}&\0\\ \0&  \widetilde{\Q}^{(k)}\esmx^T \R_{i:n,i:n}^{(k-1)}  \bsmx \I_{k-i}&\0\\ \0&\widetilde{\Z}^{(k)}\esmx.
\end{align}
Then, from \eqref{e:rink} and \eqref{e:diagk-1}, for $i=1,\ldots, k-1$,
\begin{align*}
|r^{(k)}_{ii}|
&=|r^{(k-1)}_{ii}| = \min_{\x\,\in\,\mathbb{Z}^{n-i+1}\backslash \{\0\}}\| \R^{(k-1)}_{i:n,i:n} \x\|_2 \\
%&= \min_{\x\,\in\,\mathbb{Z}^{n-i+1}\backslash \{\0\}}\| \bar{\Q}^{(k)} \R^{(k)}_{i:n,i:n}(\bar{\Z}^{(k)})^{-1} \x\|_2 \\
&=\min_{\z\,\in\,\mathbb{Z}^{n-i+1}\backslash \{\0\}}\| \R^{(k)}_{i:n,i:n}\z\|_2
\end{align*}
where  $\z=\bsmx \I_{k-i}&\0\\ \0&\widetilde{\Z}^{(k)}\esmx^{-1}\x$.
From \eqref{e:Rk}, \eqref{e:recursionk-12}, \eqref{e:zkxk} and \eqref{e:SVPk},
\begin{align}
|r^{(k)}_{kk}|&=\|\widetilde{\R}^{(k)}\e_1\| =\|(\widetilde{\Q}^{(k)})^T\R^{(k-1)}_{k:n,k:n}\widetilde{\Z}^{(k)}\e_1\| \nonumber \\
& =  \|\R^{(k-1)}_{k:n,k:n}\x^{(k)}\|
  =  \min_{\x\,\in \mathbb{Z}^{n-k+1}\setminus \{\0\}}\|\R^{(k-1)}_{k:n,k:n}\x\|_2\nonumber \\
& = \min_{\x\,\in \mathbb{Z}^{n-k+1}\setminus \{\0\}}\|  \widetilde{\R}^{(k)} (\widetilde{\Z}^{(k)})^{-1}\x \|_2  \nonumber \\
& = \min_{\z\,\in \mathbb{Z}^{n-k+1}\setminus \{\0\}}\|\R^{(k)}_{k:n,k:n}\z\|_2.
\label{e:diagkk}
\end{align}
Thus \eqref{e:diagk-1} holds when $k-1$ changes to $k$.
Then, with  $\bbR=\R^{(n-1)}$, we can conclude  \eqref{e:criteria2KZ} holds.

In the following, we introduce the process of obtaining the unimodular matrix $\widetilde{\Z}^{(k)}$ in \eqref{e:zkxk} proposed in \cite{ZhaQW12}.
(There are some other methods to find $\widetilde{\Z}^{(k)}$, see, e.g., \cite[pp.13]{New72}.)
Suppose that $\z=[p,q]^T\in \mathbb{Z}^2$ and $\gcd(p,q)=d$, then, there exist two integers $a$ and $b$ such that $ap+bq=d$.
Obviously,
\begin{align}
\label{e:U}
\U=\bmx p/d &-b\\ q/d &a  \emx
\end{align}
is unimodular and it is easy to verify that $\U^{-1} \z = d \,\e_1$.
%$$
%\U^{-1} \z = d \,\e_1.
%$$

From \eqref{e:SVPk}, we  can conclude that
$$
\gcd(x^{(k)}_1,  x^{(k)}_2, \ldots,  x^{(k)}_k)=1.
$$
After getting $\x^{(k)}$,   $\widetilde{\Z}^{(k)}$ can be obtained by applying a sequence of 2 by 2 unimodular transformations of the form
\eqref{e:U} to transform $\x^{(k)}$ to $\e_1$, i.e., $(\widetilde{\Z}^{(k)})^{-1} \x^{(k)}=\e_1$ (see \eqref{e:zkxk}).
%Specifically, suppose we want to transform $\x^{(k)}_{i+1}$ to 0 for $1$.
Specifically they  eliminate the entries of $ \x^{(k)}$ from the last one to the second one.
The resulting algorithm for finding $\widetilde{\Z}^{(k)}$ is described by Algorithm \ref{a:expansion}  and
the corresponding KZ reduction algorithm is described by Algorithm \ref{a:KZ}.

\begin{algorithm}[h!]
\caption{The Basis Expansion Algorithm in \cite{ZhaQW12}}   \label{a:expansion}
%\textbf{Input:} an upper triangular $\R \in \Rnbn$, a unimodular $\Z\in \mathbb{Z}^{n\times n}$,
%the index $k$ and $\x\,\in \mathbb{Z}^{n-k+1}$ such that:\\
%$$\x=\arg\min_{\z\,\in\,\mathbb{Z}^{n-i+1}\backslash \{\0\}}\|\R_{k:n,k:n}\z\|_2.$$
%\textbf{Output:} the updated upper triangular $\R$ with $r_{kk}=\|\R_{k:n,k:n}\x\|_2$
%%$$r_{kk}=\|\R_{k:n,k:n}\x\|_2$$
%and the updated unimodular matrix $\Z$.

\begin{algorithmic}[1]
    \FOR{$i=n-k,\dots,1$}
      \STATE find $d=\gcd(x_{i}, x_{i+1})$ and  integers $a$ and $b$ such that $ax_{i}+bx_{i+1}=d$;
      \STATE set $\U=\bmx x_{i}/d &-b\\x_{i+1}/d &a\\\emx$; \; $x_{i}=d$;
      \STATE $\Z_{1:n,i+k-1:i+k}=\Z_{1:n,i+k-1:i+k}\U$;
      \STATE $\R_{1:i+k,i+k-1:i+k}=\R_{1:i+k,i+k-1:i+k}\U$;
      \STATE find a 2 by 2 Givens rotation $\G$ such that:
      $$\G\bmx r_{i+k-1,i+k-1}\\r_{i+k,i+k-1}\\\emx=\bmx \times\\0\\\emx;$$
      \STATE $\R_{i+k-1:i+k,i+k-1:n}=\G\R_{i+k-1:i-k,i+k-1:n}$;
      %\STATE $\Q_{1:n,i+k-1:i+k}= \Q_{1:n,i+k-1:i+k} \G^T$;
    \ENDFOR
\end{algorithmic}
\end{algorithm}

\begin{algorithm}[h!]
\caption{The KZ Reduction Algorithm in \cite{ZhaQW12}}   \label{a:KZ}
%\textbf{Input:} $\A\in \mathbb{R}^{m\times n}$\\
%\textbf{Output:} a KZ reduced upper triangular $\R \in\mathbb{R}^{n\times n}$
%and the corresponding unimodular matrix $\Z\in \mathbb{Z}^{n\times n}$.

\begin{algorithmic}[1]
    \STATE computer the QR factorization of $\A$, see \eqref{e:QR};
    \STATE set $\Z=\I$;
    \FOR{$k=1$ to $n-1$}
      \STATE  solve
      $\min_{\x\,\in \mathbb{Z}^{n-k+1}\setminus \{\0\}}\|\R_{k:n,k:n}\x\|_2^2$
      by the LLL-aided Schnorr-Euchner search strategy;
%      use the LLL-aided Schnorr-Euchner search strategy to find a vector $\x\in\mathbb{Z}^{n-k+1}$
%      such that $\R_{k:n,k:n}\x$ is a shortest nonzero point in the lattice generated by $\R_{k:n,k:n}$;
      \STATE apply Algorithm \ref{a:expansion} to update $\R$ and $\Z$;
    \ENDFOR
    \STATE perform size reductions on $\R$ and update $\Z$
\end{algorithmic}
\end{algorithm}

Here we make a remark.  Algorithm \ref{a:KZ} does not show how to form and update $\Q$, as
it may not be needed in applications.
If an application indeed needs $\Q$, then we can obtain it by the QR factorization of $\A\Z$
after obtaining $\Z$. This would be more efficient.

\subsection{Proposed KZ Reduction Algorithm }
In this subsection, we modify Algorithms \ref{a:expansion} and \ref{a:KZ}  to get a new KZ reduction algorithm,
which can be much faster and more numerically reliable.

First, we make an observation on Algorithm \ref{a:KZ} and make a simple modification.
At step $k$, if $\x^{(k)}=\pm\,\e_1$  (see \eqref{e:SVPk}),
%\begin{align}
%\label{e:sv}
%\|\R^{(k-1)}_{k:n,k:n}\x^{(k)}\|=r^{(k-1)}_{kk},
%\end{align}
then, obviously, the basis expansion algorithm, i.e., Algorithm \ref{a:expansion} is not needed
and we can move to step $k+1$.
Later we will come back to this issue again.

In the following, we will  make some major modifications.
But before doing it,  we  introduce the following basic fact, which can be found in the literature:
%\begin{lemma} \label{l:gcd}
For any two integers $p$ and $q$, the time complexity of finding two integers $a$ and $b$ such that $ap+bq=d\equiv \gcd(p,q)$
by  the extended Euclid algorithm is bounded by $\bigO(\log_2(\min\{|p|,|q|\}))$ if  fixed precision is used.
%\end{lemma}

In Algorithm \ref{a:KZ}, after finding $\x^{(k)}$ (see \eqref{e:SVPk}), Algorithm \ref{a:expansion} is used to expand
$\R^{(k-1)}_{k:n,k:n}\x^{(k)}$ to a basis for the lattice $\{\R^{(k-1)}_{k:n,k:n}\x: \x\in \mathbb{Z}^{n-k+1}\}$.
There are some serious drawbacks with this approach.
Sometimes, especially when $\A$ is ill-conditioned, some of the entries of $\x^{(k)}$ may be very large
such that they are beyond the range of consecutive integers in a floating point system (i.e., integer overflow occurs),
very likely resulting in wrong results.
Even if integer overflow does not occur in storing $\x^{(k)}$, large $\x^{(k)}$ may still cause problems.
One problem is that the computational time of the extended Euclid algorithm will be long
according to its complexity result we just mentioned before.
The second problem is that  updating $\Z$ and  $\R$ in lines 4 and 5 of Algorithm \ref{a:expansion}
may cause numerical issues.
Large $x_i$ and $x_{i+1}$ are likely to produce large elements in $\U$.
As a result, integer overflow may occur in updating $\Z$,
and large rounding errors are likely to occur in updating  $\R$.
Finally, $\R$ is likely to become more ill-conditioned after the updating,
making the search process for solving SVPs in later steps expensive.

In order to deal with the large $\x^{(k)}$ issue,
we look at line 4 in  Algorithm \ref{a:KZ}, which uses the LLL-aided Schnorr-Euchner search strategy to solve the SVP.
Specifically at step $k$, to solve \eqref{e:SVPk}, the LLL reduction algorithm is applied to $\R^{(k-1)}_{k:n,k:n}$:
\beq
\label{e:QRZk2}
(\widehat{\Q}^{(k)})^T\R^{(k-1)}_{k:n,k:n}\widehat{\Z}^{(k)}=\widehat{\R}^{(k-1)}
\eeq
where $\widehat{\Q}^{(k)}\in \mathbb{R}^{(n-k+1)\times(n-k+1)}$ is orthogonal, $\widehat{\Z}^{(k)}\in \mathbb{Z}^{(n-k+1)\times(n-k+1)}$
is unimodular and $\widehat{\R}^{(k-1)}$ is LLL-reduced.
Then, one solves the reduced SVP by the Schnorr-Euchner search strategy:
\begin{align}
\label{e:SVPk2}
\z^{(k)}=\arg \min_{\z\,\in \mathbb{Z}^{n-k+1}\setminus \{\0\}}\|\widehat{\R}^{(k-1)}\z\|_2^2.
\end{align}
The solution of the original SVP is $\x^{(k)}=\widehat{\Z}^{(k)}\z^{(k)}$.

Instead of expanding $\R^{(k-1)}_{k:n,k:n}\x^{(k)}$ as done in Algorithm \ref{a:KZ},
we propose to expand $\widehat{\R}^{(k-1)}\z^{(k)}$
to a basis for the lattice $\{\widehat{\R}^{(k-1)}\z: \z\in \mathbb{Z}^{n-k+1}\}$.
Thus, before doing the expansion, we update $\Q^{(k)}, \R^{(k)}$ and $\Z^{(k)}$ by using the LLL reduction
\eqref{e:QRZk2}:
\begin{align}
\label{e:Qk2}
\check{\Q}^{(k)}&=\Q^{(k-1)}\bsmx \I_{k-1}&\0\\ \0&\widehat{\Q}^{(k)}\esmx,\\
\label{e:Rk2}
\check{\R}^{(k)}&=\bsmx \R^{(k-1)}_{1:k-1, 1:k-1}&\R^{(k-1)}_{1:k-1, k:n}\widehat{\Z}^{(k)}\\ \0&\widehat{\R}^{(k-1)}\esmx,\\
\check{\Z}^{(k)}&=\Z^{(k-1)}\bsmx \I_{k-1}&\0\\ \0&\widehat{\Z}^{(k)}\esmx.
\label{e:Zk2}
\end{align}
Now we do expansion.
We construct  a  unimodular matrix $\widetilde{\Z}^{(k)}\in \Zbb^{(n-k+1)\times (n-k+1)}$ whose first column is $\z^{(k)}$,
and find  an orthogonal matrix $\widetilde{\Q}^{(k)}$ to bring $\widehat{\R}^{(k-1)}\widetilde{\Z}^{(k)}$
back to an upper triangular matrix $\widetilde{\R}^{(k)}$ (cf.\ \eqref{e:recursionk-12}):
\beq \label{e:recursionk-13}
(\widetilde{\Q}^{(k)})^T\widehat{\R}^{(k-1)}\widetilde{\Z}^{(k)}=\widetilde{\R}^{(k)}.
\eeq
Then, we update $\check{\Q}^{(k)}$, $\check{\R}^{(k)}$ and $\check{\Z}^{(k)}$ as follows
 (cf.\ \eqref{e:Qk}--\eqref{e:Zk}):
\begin{align}
\label{e:Qk3}
\Q^{(k)}&=\check{\Q}^{(k-1)}\bsmx \I_{k-1}&\0\\ \0&\widetilde{\Q}^{(k)} \esmx,\\
\label{e:Rk3}
\R^{(k)}&=\bsmx \check{\R}^{(k-1)}_{1:k-1, 1:k-1}&\check{\R}^{(k-1)}_{1:k-1, k:n}\widetilde{\Z}^{(k)} \\ \0&\widetilde{\R}^{(k)}\esmx,\\
\Z^{(k)}&=\check{\Z}^{(k-1)}\bsmx \I_{k-1}&\0\\ \0&\widetilde{\Z}^{(k)} \esmx
\label{e:Zk3}
\end{align}
and we obtain the QRZ factorization of $\R$ in the same form as \eqref{e:recursionk} at step $k$.

Unlike $\x^{(k)}$ in \eqref{e:SVPk}, which can be arbitrarily large, $\z^{(k)}$ in \eqref{e:SVPk2} can be bounded.
Actually by using the LLL reduction properties and the fact that
$$
\|\widehat{\R}^{(k-1)}\z^{(k)}\|_2 \leq \|\widehat{\R}^{(k-1)}\e_1\|_2=|\widehat{r}_{11}^{(k-1)}|
$$
we can show the following result:
\begin{theorem}
\label{t:zk}
For $1\leq i\leq n-k+1$, the $i$-th entry of $\z^{(k)} \in \mathbb{Z}^{n-k+1}$ (see \eqref{e:SVPk2}) satisfies
\begin{align}
\label{e:zk}
|z^{(k)}_i |  \leq\big(\frac{4}{4\delta-1}\big)^{(n-k)/2}2^{n-k+1-i}
\end{align}
where $\delta$ is the parameter in the LLL reduction (see \eqref{e:criteria2}).
\end{theorem}
%{\bf Proof}. $\widehat{\R}^{(k-1)}$ is LLL reduced, so by \eqref{e:criteria2}, for
%$1\leq i\leq n-k+1$, we have:
%\begin{align}
%\label{e:ratiodiag}
%|\frac{\widehat{r}_{11}^{(k-1)}}{\widehat{r}_{ii}^{(k-1)}}|\leq \big(\frac{4}{4\delta-1}\big)^{(i-1)/2}
%\end{align}
%and for $i<l\leq n-k+1$, we have:
%\begin{align}
%\label{e:nondiag}
%|\widehat{r}_{il}^{(k-1)}|\leq |\widehat{r}_{ii}^{(k-1)}|.
%\end{align}
%
%We prove \eqref{e:zk} by induction on decreasing $i$.
%
%Suppose $i=n-k+1$, since
%$$
%\|\widehat{\R}^{(k-1)}\z^{(k)}\|_2 \leq |\widehat{r}_{11}^{(k-1)}|
%$$
%we have
%$$
%|\widehat{r}^{(k-1)}_{n-k+1, n-k+1}\z^{(k)}_{n-k+1}| \leq |\widehat{r}_{11}^{(k-1)}|.
%$$
%Then, by \eqref{e:ratiodiag}, \eqref{e:zk} obviously holds for $i=n-k+1$.
%
%Suppose that \eqref{e:zk} obviously holds for $i\geq j+1$ ($1\leq j\leq n-k$). In the following, we show \eqref{e:zk} also holds for $i=j$.
%Since
%$$
%\|\widehat{\R}^{(k-1)}\z^{(k)}\|_2 \leq |\widehat{r}_{11}^{(k-1)}|
%$$
%we have
%$$
%|\widehat{r}^{(k-1)}_{jj}||\z^{(k)}_{j}+\sum_{l=j+1}^{n-k+1}\frac{\widehat{r}^{(k-1)}_{jl}}{\widehat{r}^{(k-1)}_{jj}}\z^{(k)}_{l}| \leq |\widehat{r}_{11}^{(k-1)}|.
%$$
%By \eqref{e:ratiodiag}, we have
%$$
%|\z^{(k)}_{j}+\sum_{l=j+1}^{n-k+1}\frac{\widehat{r}^{(k-1)}_{jl}}{\widehat{r}^{(k-1)}_{jj}}\z^{(k)}_{l}| \leq (\frac{4}{4\delta-1}\big)^{(j-1)/2}\leq (\frac{4}{4\delta-1}\big)^{(n-k)/2}.
%$$
%Then, by \eqref{e:nondiag}, we have
%$$
%|\z^{(k)}_{j}|\leq (\frac{4}{4\delta-1}\big)^{(n-k)/2}+\sum_{l=j+1}^{n-k+1}|\z^{(k)}_{l}|.
%$$
%Then, by the induction assumption, it is not hard to show that \eqref{e:zk} also holds for $i=j$.\ \ $\Box$

Because of the limitation of space, we omit its proof.

Now we discuss the benefits of the modification.
First, since $\widehat{\R}^{(k-1)}$ is LLL reduced, it has a very good chance,
especially when $\R$ is  well-conditioned and $n$ is small (say, smaller than 30),
that $\z^{(k)}=\pm \,\e_1$ (see \eqref{e:SVPk2}).
This was observed in our simulations.
As we stated before, the basis expansion is not needed in this case and we can move to next step.
Second,  the entries of $\z^{(k)}$ are bounded according to Theorem \ref{t:zk},
but the  entries of $\x^{(k)}$ are not.
Our simulations indicated that the former are smaller  or much smaller than the latter.
Thus, the serious problems with using $\x^{(k)}$ for basis expansion mentioned before
can be significantly mitigated by using $\z^{(k)}$ instead.

To  further reduce the computational cost,
we look at the basis expansion process at step $k$ of Algorithm 2.
After  $\z^{(k)}$ is obtained,  Algorithm 1 is used to find a sequence of 2 by 2 unimodular matrices
in the form of \eqref{e:U} to eliminate its entries form the last one to the second one.
We noticed in our simulations that often $\z^{(k)}$ has a lot of zeros and we would like to explore this
to make the basis expansion process more efficient.
Specifically, if $\z=[p,q]^T\in \mathbb{Z}^2$ with $q=0$, then $\mbox{gcd}(p,q)=p$,
and $\U=\I_2$ in \eqref{e:U}. Thus, in this case we do not need to do anything
and move to eliminate the next element in  $\z^{(k)}$.
%If $p=0$ and $q\neq 0$, we interchange the two entries, i.e., %
%In the $i-$th step of construct $\widetilde{\Z}^{(k)}$, if the update $z^{(k)}_{n-k+2-i}$ is 0,
%then we set $i=i-1$ and move to the next iteration.
%Otherwise, if $z^{(k)}_{n-k+1-i}=0$, then by \eqref{e:U},  we can set
%$\U=\bsmx 0 &1\\1 &0 \esmx$.

% we make the following modification.
%Note that, during the process of the Algorithm \ref{a:expansion}, for any fixed $i$ (see line 1), if $x_{i+1}=0$, lines 2-7
%do not need to be performed. In fact, we only need to set $i=i-1$ and move to the next recursion.
%Let $\bar{\z}^{(k)}$ have $s$ zero entries, then by the above analysis, we only need to expand
%$\check{\R}^{(k-1)}_{k:n-s,k:n-s}\bar{\z}^{(k)}_{1:n-k+1-s}$ to a basis for the lattice
%$\{\check{\R}^{(k-1)}_{k:n-s,k:n-s}\x: \x\in \mathbb{Z}^{n-k+1-s}\}$.

Now we can describe the modified  KZ reduction algorithm in Algorithm \ref{a:mKZ}.

\begin{algorithm}[h!]
\caption{Modified KZ Reduction Algorithm}   \label{a:mKZ}
%\textbf{Input:} $\A\in \mathbb{R}^{m\times n}$\\
%\textbf{Output:} a KZ reduced upper triangular $\R \in\mathbb{R}^{n\times n}$
%and the corresponding unimodular matrix $\Z\in \mathbb{Z}^{n\times n}$.\\
\begin{algorithmic}[1]
    \STATE computer the QR factorization of $\A$, see \eqref{e:QR};
    \STATE set $\Z=\I, k=1$;
   % \STATE set $k=1$;
    \WHILE{$k<n$}
      \STATE compute the LLL reduction of $\R_{k:n,k:n}$ (see \eqref{e:QRZk2}) and update $\R, \Z$ (see \eqref{e:Rk2}-\eqref{e:Zk2});
      \STATE solve
      $\min_{\z\,\in \mathbb{Z}^{n-k+1}\setminus \{\0\}}\|\R_{k:n,k:n}\z\|_2^2$
      by the Schnorr-Euchner search strategy to get the solution $\z$;
      \IF{$\z=\pm\,\e_1$}
      \STATE $k=k+1$;
      \ELSE
      \STATE $i=n-k$;
         \WHILE{$i\geq1$}
           \IF{$z_{i+1}\neq0$}
              \STATE  perform lines 2-7 of Algorithm \ref{a:expansion} (where $x_i$ and $x_{i+1}$
               are replaced by $z_i$ and $z_{i+1}$);
           \ENDIF
         \STATE $i=i-1$;
         \ENDWHILE
         \STATE $k=k+1$;
      \ENDIF
    \ENDWHILE
    \STATE perform size reductions on $\R$ and update $\Z$.
\end{algorithmic}
\end{algorithm}

%Algorithm \ref{a:mKZ} does not show how to form and update $\Q$. If it is needed for some application, we can use
%the method similar to that stated in Remark \ref{r:formQ} to form it.

\section{Numerical tests} \label{s:sim}

In this section, we compare the performance of the proposed KZ algorithm Algorithm \ref{a:mKZ} with Algorithm \ref{a:KZ}.
All the numerical tests  were done by \textsc{Matlab} 14b on a desktop computer with Intel(R)
Xeon(R) CPU W3530 @ 2.80GHz$\times4$.
The \textsc{Matlab} code for Algorithm \ref{a:KZ} was provided by Dr.\ Wen Zhang, one of the authors of \cite{ZhaQW12}.
The parameter $\delta$  in the LLL reduction was chosen to be 1.

We first give an example to show that Algorithm \ref{a:KZ} may not even give a LLL reduced matrix (for $\delta=1$),
while Algorithm \ref{a:mKZ} does.

%\begin{example}
\textit{Example}.
Let
%\footnotesize
$$
%\A=\bmx
%29.0515&-6.2868&-13.824&35.8878&56.9155\\
%0&3.1479&-0.3457&2.2674&4.8752\\
%0&0&0.2320&-0.3432&-0.4627\\
%0&0&0&0.0102&0.0335\\
%0&0&0&0&0.0035
%\emx
\A\!=\!\bmxc{rrrrr}
10.6347&-66.2715&9.3046&17.5349&24.9625\\
0&8.6759&-4.7536&-3.9379&-2.3318\\
0&0&0.3876&0.1296&-0.2879\\
0&0&0&0.0133&-0.0082\\
0&0&0&0&0.0015
\emxc.
$$

Applying Algorithm \ref{a:KZ} gives
%{\footnotesize
$$
\R=\bmxc{rrrrr}
-0.2256&-0.0792&0.0125&0&0\\
0&0.2148&-0.0728&-0.0029&-0.0012\\
0&0&0.2145&0.0527&-0.0211\\
0&0&0&-0.1103&0.0306\\
0&0&0&0&0.6221
\emxc.
$$
%}

It is easy to check that $\R$ is not LLL reduced (for $\delta=1$) since ${r}_{33}^2>{r}_{34}^2+{r}_{44}^2$.
Moreover, the matrix $\Z$ obtained by Algorithm \ref{a:KZ} is not unimodular since its determinant is $-3244032$,
which was precisely calculated by Maple.
The reason for this is that $\A$ is  ill conditioned
(its condition number in the 2-norm is about $1.0\times 10^5$) and some of the entries of $\x^{(k)}$ (see \eqref{e:SVPk}) are too large,
causing  severe inaccuracy in updating $\R$ and integer overflow in updating $\Z$ (see lines 4-5 in Algorithm \ref{a:expansion}).
In fact,
\begin{align*}
\x^{(1)}&=\bmx
-47,&-27,&-21,&-14,&-34
\emx^T;\\
\x^{(2)}&=\bmx
-48029,&-27593,&2145,&345
\emx^T; \\
\x^{(3)}&=\bmx
-2767925153,&432235,&40
\emx^T;\\
\x^{(4)}&=\bmx
691989751,&2
\emx^T.
\end{align*}
The condition numbers in the 2-norm of $\R(k\!:\!5,k\!:\!5)$
obtained at the end of step $k=1,2,3,4$ of Algorithm \ref{a:KZ} are respectively
$2.9\times10^8, 1.5\times10^{15},6.2 \times10^{18}$ and $1.1\times 10$.
A question one may raise is  that if $\A$ is updated by the unimodular matrices
produced in the process (i.e., $\Z$ is not explicitly formed) is $\A\Z$ LLL reduced?
We found it is still not by looking at the R-factor of the QR factorization of $\A\Z$.

Applying Algorithm \ref{a:mKZ} to $\A$ gives
$$
%{\footnotesize
{\R}=\bmxc{rrrrr}
-0.2256&0.0792&-0.0126&0.0028&-0.0621\\
0&-0.2148&0.0728&-0.0084&0.0930\\
0&0&0.2145&0.0292&-0.0029\\
0&0&0&-0.2320&0.0731\\
0&0&0&0&-0.2959
\emxc.
%}
$$
Although we cannot verify if $\R$ is KZ reduced, we can verify that indeed it is LLL reduced.
\noindent All of the solutions of the four SVPs are $\e_1$ (note that the dimensions are different).
Thus, no basis expansion is needed.
The condition numbers in the 2-norm of $\R(k\!:\!5,k\!:\!5)$ obtained at the end of step
$k=1,2,3,4$ of Algorithm \ref{a:mKZ} are respectively
$2.1, 1.9, 1.6$ and $1.4$.

%\end{example}

%Note that we found many examples in show that the Algorithm \ref{a:KZ} can not return a
%correct reduced matrix after applying it on a not very small dimensional matrix which
%is ill-conditioned.

Now we consider two more general cases for comparing the efficiency of the two algorithms:
\begin{itemize}
\item  Case 1. $\A=\text{randn}(n,n)$,
where $\text{randn}(n,n)$ is a \textsc{Matlab} built-in function
to  generate a random $n\times n$ matrix, whose entries follow the normal distribution ${\cal N}(0,1)$.

\item  Case 2. $\A=\U\D\V^T$, $\U,\V$ are random orthogonal matrices obtained by the QR factorization of random matrices generated by
$\text{randn}(n,n)$ and $\D$ is a $n\times n$ diagonal matrix with $d_{ii}=10^{3(n/2-i)/(n-1)}$.

\end{itemize}

\begin{figure}[!htbp]
\centering
\includegraphics[width=3.2in]{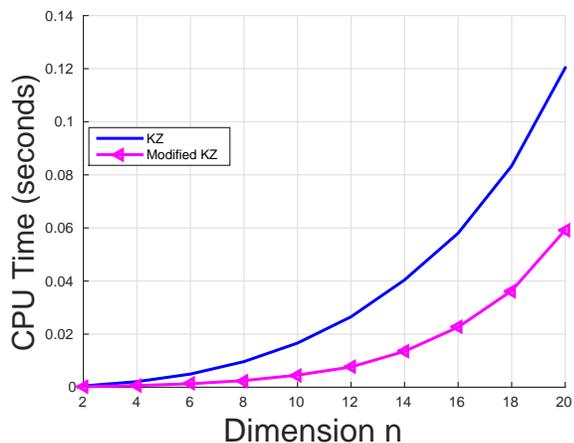}
\caption{Average CPU time   versus $\n$ for Case 1} \label{fig:CPUT1}
\end{figure}

\begin{figure}[!htbp]
\centering
\includegraphics[width=3.2in]{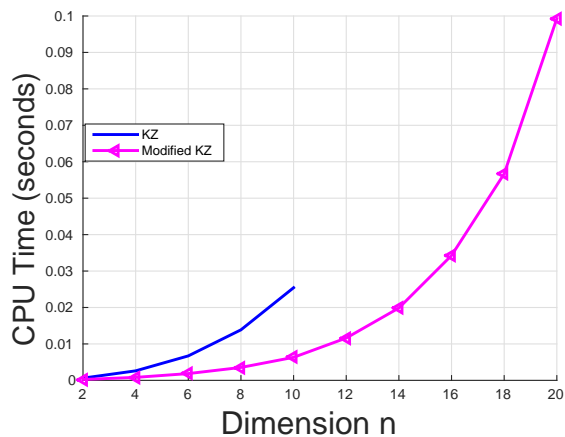}
\caption{Average CPU time   versus $\n$ for Case 2} \label{fig:CPUT2}
\end{figure}

In the numerical tests for each case for a fixed $n$ we gave 200 runs to generate 200 different $\A$'s.
Figures  \ref{fig:CPUT1} and \ref{fig:CPUT2} display  the average CPU time over 200 runs versus $n=2:2:20$
for Cases 1 and 2, respectively. In both figures, ``KZ'' and ``Modified KZ'' refer to Algorithms \ref{a:KZ} and \ref{a:mKZ},
respectively.

Figure \ref{fig:CPUT2} gives the results for only $n=2:2:10$. This is because when $n\geq12$,
Algorithm 2 often did not terminate within ten hours.  %and the results are usually not correct.

In Case 1, sometimes Algorithm \ref{a:KZ} did  not terminate within a half hour
and we just ignored this instance and gave one more run.
The number of such instances was much smaller than that for Case 2.

From Figures  \ref{fig:CPUT1} and \ref{fig:CPUT2}, we can see that Algorithm \ref{a:mKZ} is faster than
Algorithm \ref{a:KZ} for Case 1 and  much faster  for Case 2.
Also, when we ran Algorithm \ref{a:KZ} we got a warning message
"Warning: Inputs contain values larger than the largest consecutive flint. Result may be inaccurate" several times,
for both Cases 1 and 2 in the tests. But this did not happen to Algorithm \ref{a:mKZ}.
Thus Algorithm \ref{a:mKZ} is more numerically reliable.

\section{Summary and comment} \label{s:sum}
In this paper, we modified the KZ reduction algorithm proposed by  Zhang et al. in \cite{ZhaQW12}.
The resulting algorithm  can be much faster and more numerically reliable.

The modified basis expansion strategy proposed in this paper can be applied in designing algorithms for the Minkowski reduction
(see, e.g., \cite{ZhaQW12}) and the block KZ reduction (see \cite{Sch87} and \cite{CheN11}).

\bibliographystyle{IEEEtran}
\bibliography{ref}

\end{document}

%% file: micros.tex
\usepackage{amsmath}
\usepackage{amsfonts}
\usepackage{amssymb}
\usepackage{algorithm}
\usepackage{algorithmic}
\usepackage{graphicx}
\usepackage{verbatim}

\newtheorem{theorem}{Theorem}

\newtheorem{example}{Example}

\newcommand{\beq}{\begin{equation}}
\newcommand{\eeq}{\end{equation}}
\newcommand{\beqnn}{\begin{equation*}}
\newcommand{\eeqnn}{\end{equation*}}
\newcommand{\beqy}{\begin{eqnarray}}
\newcommand{\eeqy}{\end{eqnarray}}
\newcommand{\beqynn}{\begin{eqnarray*}}
\newcommand{\eeqynn}{\end{eqnarray*}}
\newcommand{\bit}{\begin{itemize}}
\newcommand{\eit}{\end{itemize}}
\newcommand{\ben}{\begin{enumerate}}
\newcommand{\een}{\end{enumerate}}
\newcommand{\bex}{\begin{example}}
\newcommand{\eex}{\end{example}}

\newcommand{\balg}[1]{\begin{algorithm} \caption{#1}}
\newcommand{\ealg}{\end{algorithm}}

\newcommand{\balgc}{\begin{algorithmic}[1]}
\newcommand{\ealgc}{\end{algorithmic}}

\newcommand{\bary}{\begin{array}}
\newcommand{\eary}{\end{array}}
\newcommand{\bmx}{\begin{bmatrix}}
\newcommand{\emx}{\end{bmatrix}}
\newcommand{\bsmx}{\left[\begin{smallmatrix}}
\newcommand{\esmx}{\end{smallmatrix}\right]}
\newcommand{\bmxc}[1]{\left[\begin{array}{@{}#1@{}}}
\newcommand{\emxc}{\end{array}\right]}
\newcommand{\bcn}{\begin{center}}
\newcommand{\ecn}{\end{center}}

\newcommand{\Rbb}{{\mathbb{R}}}
\newcommand{\Zbb}{{\mathbb{Z}}}

\newcommand{\bigO}{{\mathcal{O}}}

\newcommand{\Rnbn}{\Rbb^{n \times n}}

\newcommand{\Rmbm}{\Rbb^{m \times m}}

\newcommand{\A}{\boldsymbol{A}}

\newcommand{\D}{\boldsymbol{D}}

\newcommand{\G}{\boldsymbol{G}}

\newcommand{\I}{\boldsymbol{I}}

\renewcommand{\P}{\boldsymbol{P}}
\newcommand{\Q}{\boldsymbol{Q}}
\newcommand{\R}{\boldsymbol{R}}

\newcommand{\U}{\boldsymbol{U}}
\newcommand{\V}{\boldsymbol{V}}

\newcommand{\Z}{\boldsymbol{Z}}

\newcommand{\e}{\boldsymbol{e}}

\newcommand{\n}{\boldsymbol{n}}

\newcommand{\x}{{\boldsymbol{x}}}
\newcommand{\y}{{\boldsymbol{y}}}
\newcommand{\z}{\boldsymbol{z}}
\newcommand{\0}{{\boldsymbol{0}}}

\newcommand{\br}{{\bar{r}}}
\newcommand{\bbQ}{{\bar{\Q}}}
\newcommand{\bbR}{{\bar{\R}}}

%% file: KZ6.bbl
\begin{thebibliography}{10}
\providecommand{\url}[1]{#1}
\csname url@samestyle\endcsname
\providecommand{\newblock}{\relax}
\providecommand{\bibinfo}[2]{#2}
\providecommand{\BIBentrySTDinterwordspacing}{\spaceskip=0pt\relax}
\providecommand{\BIBentryALTinterwordstretchfactor}{4}
\providecommand{\BIBentryALTinterwordspacing}{\spaceskip=\fontdimen2\font plus
\BIBentryALTinterwordstretchfactor\fontdimen3\font minus
  \fontdimen4\font\relax}
\providecommand{\BIBforeignlanguage}[2]{{%
\expandafter\ifx\csname l@#1\endcsname\relax
\typeout{** WARNING: IEEEtran.bst: No hyphenation pattern has been}%
\typeout{** loaded for the language `#1'. Using the pattern for}%
\typeout{** the default language instead.}%
\else
\language=\csname l@#1\endcsname
\fi
#2}}
\providecommand{\BIBdecl}{\relax}
\BIBdecl

\bibitem{AgrEVZ02}
E.~Agrell, T.~Eriksson, A.~Vardy, and K.~Zeger, ``Closest point search in
  lattices,'' \emph{IEEE Transactions on Information Theory}, vol.~48, no.~8,
  pp. 2201--2214, 2002.

\bibitem{KZ73}
A.~Korkine and G.~Zolotareff, ``Sur les formes quadratiques,''
  \emph{Mathematische Annalen}, vol.~6, no.~3, pp. 366--389, 1873.

\bibitem{Min96}
H.~Minkowski, ``Geometrie der zahlen (2 vol.),'' \emph{Teubner, Leipzig}, vol.
  1910, 1896.

\bibitem{LenLL82}
A.~Lenstra, H.~Lenstra, and L.~Lov{\'a}sz, ``Factoring polynomials with
  rational coefficients,'' \emph{Mathematische Annalen}, vol. 261, no.~4, pp.
  515--534, 1982.

\bibitem{Sey93}
M.~Seysen, ``Simultaneous reduction of a lattice basis and its reciprocal
  basis,'' \emph{Combinatorica}, vol.~13, no.~3, pp. 363--376, 1993.

\bibitem{HanPS11}
G.~Hanrot, X.~Xavier~Pujol, and D.~Stehl{\'e}, ``Algorithms for the shortest
  and closest lattice vector problems,'' in \emph{IWCC'11 Proceedings of the
  Third international conference on Coding and cryptology}, 2011, pp. 159--190.

\bibitem{WubSJM11}
D.~W\"{u}bben, D.~seethaler, J.~Jald{\'e}n, and G.~Matz, ``Lattice reduction: A
  survey with applications in wireless communications,'' \emph{IEEE
  Transactions on Magazine}, vol.~28, no.~3, pp. 79--91, 2011.

\bibitem{Teu96}
P.~J.~G. Teunissen, \emph{{GPS} carrier phase ambiguity fixing concepts. In
  Kleusberg A and Teunissen, P. J. G, editors, {GPS} for Geodesy, pp.
  317-388}.\hskip 1em plus 0.5em minus 0.4em\relax Springer, Heidelberg.

\bibitem{ChaWX13}
X.-W. Chang, J.~Wen, and X.~Xie, ``Effects of the {LLL} reduction on the
  success probability of the {B}abai point and on the complexity of sphere
  decoding,'' \emph{IEEE Transactions on Information Theory}, vol.~59, no.~8,
  pp. 4915--4926, 2013.

\bibitem{Hel85}
B.~Helfrich, ``Algorithms to construct minkowski reduced and hermite reduced
  lattice bases,'' \emph{Theoretical Computer Science}, vol.~41, no.~8, pp.
  125--139, 1985.

\bibitem{Kan87}
R.~Kannan, ``Minkowski's convex body theorem and integer programming,''
  \emph{Mathematics of operations research}, vol.~12, no.~3, pp. 415--440,
  1987.

\bibitem{Sch87}
C.~P. Schnorr, ``A hierarchy of polynomial time lattice basis reduction
  algorithms,'' \emph{Theoretical Computer Science}, vol.~53, pp. 201--224,
  1987.

\bibitem{ZhaQW12}
W.~Zhang, S.~Qiao, and Y.~Wei, ``{HKZ} and {M}inkowski reduction algorithms for
  lattice-reduction-aided {MIMO} detection,'' \emph{IEEE Transactions on Singal
  Processing}, vol.~60, no.~11, pp. 5963--5976, 2012.

\bibitem{SchE94}
C.~Schnorr and M.~Euchner, ``Lattice basis reduction: improved practical
  algorithms and solving subset sum problems,'' \emph{Mathematical
  Programming}, vol.~66, pp. 181--191, 1994.

\bibitem{New72}
M.~Newman, \emph{Integral Matrices}.\hskip 1em plus 0.5em minus 0.4em\relax
  Academic Press, New York and London.

\bibitem{CheN11}
Y.~Chen and P.~Q. Nguyen, ``{BKZ} 2.0: Better lattice security estimates,'' in
  \emph{Advances in Cryptology -- Proceedings of ASIACRYPT '11}, ser. LNCS,
  vol. 7073.\hskip 1em plus 0.5em minus 0.4em\relax Springer, 2011, pp. 1--20.

\end{thebibliography}
